\documentclass[letterpaper,aps,prc,superscriptaddress,showpacs,nofootinbib,floatfix,twocolumn]{revtex4-1}

\usepackage{hyperref}
\usepackage{color}
\usepackage{graphicx}	
\usepackage{chngpage}
\graphicspath{%
  {./},%
  {./figures/}%
}

\usepackage{xspace}	

\begin{document}

\title{Heavy Ion Initial Conditions and Correlations Between Higher Moments in the Spatial Anisotropy}

\author{J. L. Nagle} \affiliation{University of Colorado at Boulder}
\email{jamie.nagle@colorado.edu}

\author{M. P. McCumber} \affiliation{University of Colorado at Boulder}
\email{michael.mccumber@colorado.edu}

\date{\today}

\begin{abstract}

  Fluctuations in the initial conditions for relativistic heavy ion
  collisions are proving to be crucial to understanding final state
  flow and jet quenching observables.  The initial geometry has been
  parametrized in terms of moments in the spatial anisotropy
  (i.e. $\epsilon_{2}, \epsilon_{3}, \epsilon_{4}, \epsilon_{5}...$), and it has
  been stated in multiple published articles that the vector directions
  of odd moments are uncorrelated with the even moments and the
  reaction place angle.  In this article, we demonstrate that
  this is incorrect and that a substantial non-zero correlation exists between
  the even and odd moments in peripheral Au+Au collisions.  These correlations
  persist for all centralities, though at a very small level for the 0-55\% most
  central collisions.

\end{abstract}

\maketitle

One proposal for modeling the initial geometry fluctuations in
relativistic heavy ion collisions is utilizing a Monte Carlo Glauber
calculation~\cite{Miller:2007ri}.  Using this model and the initial
transverse positions of the struck nucleons, referred to as
participants, one can calculate the participant eccentricity
$\epsilon_{2}$ and higher moments~\cite{Alver:2010gr}. In~\cite{Alver:2010gr}, these moments are defined by:
\begin{equation}
  \epsilon_{n} = \frac{\sqrt{\left< r^{2}cos(n\phi_{part})\right>^{2} + \left< r^{2} sin(n\phi_{part})\right>^{2}}}{\left< r^{2} \right>}
\end{equation}
where $n$ is the $n$th moment of the spatial anisotropy calculated
relative to the mean position. The major axis
associated with the $n$th moment is defined by:
\begin{equation}
  \psi_{n} = \frac{atan2(\left< r^{2}sin(n\phi_{part})\right>, \left< r^{2} cos(3\phi_{part}) \right> )}{n}
\end{equation}

\begin{figure}[b]
  \centering
  \includegraphics[width=0.95\linewidth]{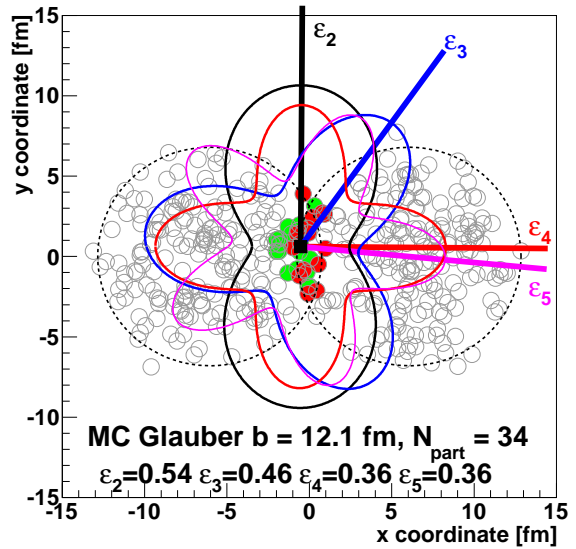}
  \caption{\label{fig_display} (Color online) Monte Carlo Glauber event display for
    a sample Au+Au collision. The grey circles represent the
    positions of all nucleons. The green (red) circles are participant
    nucleons from the left (right) nucleus. The vector directions for
    the $n=2-5$ and the spatial anisotropy pattern they represent are
    overlaid. These are centered on the mean position as indicated by a
    black square.}
\end{figure}

We show an example event display in Figure~\ref{fig_display} that
includes the positions of the participant nucleons, and a
visualization of the $\epsilon_{2},~\epsilon_{3},~\epsilon_{4}$, and
$\epsilon_{5}$ moments.  We have drawn the vector direction of each
along the long-axis of the associated moment.  The $n$th
moment has an $n$-multiplet of directions that are equally valid,
separated by $2\pi/n$.

\begin{figure}[tb]
  \centering
  \includegraphics[width=1.0\linewidth]{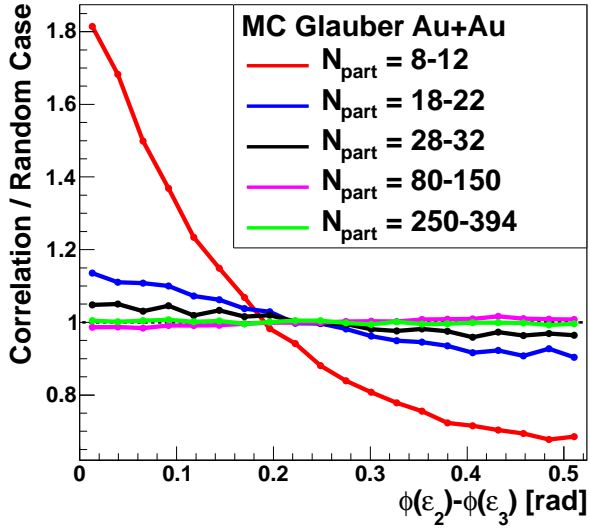}
  \caption{\label{fig_delta23} (Color online) Monte Carlo Glauber Au+Au
    distribution for relative angular difference $(\epsilon_{2} - \epsilon_{3})$
    [radians] for different selections in $N_{part}$. A flat
    distribution at one from angular separations 0.0-$\pi/6$ indicates
    totally uncorrelated quantities.}
\end{figure}

\section{Previous Results}

In~\cite{Alver:2010gr}, the authors state that the minor axis of
triangularity (i.e.~$\epsilon_{3}$) is found to be uncorrelated with
the minor axis of eccentricity (i.e.~$\epsilon_{2}$) in Monte Carlo
Glauber calculations.  This conclusion is repeated
in~\cite{Agakishiev:2010ur} where it is stated that the orientation of triangular
overlap shape due to fluctuations is random relative to the
event-plane direction as determined by the elliptic anisotropy. This
conclusion is then generalized in~\cite{Lacey:2010hw} where these authors
state that the angles for the odd and even harmonics are uncorrelated.

There have been many initial studies for how this spatial anisotropy
translates into the final momentum space distribution of particles
(e.g.~\cite{Alver:2010gr,Qin:2010pf,Alver:2010dn,Holopainen:2010gz,Teaney:2010vd,Schenke:2010rr}).
The fact that the angular orientations are uncorrelated between odd
and even moments has an important impact on the methodology for
experiments to determine these momentum anisotropy moments $v_{n}$ and
for two-particle correlation measurements, with relevance for jet
quenching observables.

\section{Methodology}

We set out to confirm the findings of the above papers using the Monte
Carlo Glauber framework.  We have utilized the standard PHOBOS Monte
Carlo Glauber code~\cite{Alver:2008aq} with Woods-Saxon
parameters and settings ($R_{0} = 6.38$ fm, $a = 0.535$ fm, $d_{min} =
0$ fm). We show the angular distribution between $\epsilon_{2}$ and $\epsilon_{3}$ for a set of Au+Au number of
participant ($N_{part}$) selections in Figure~\ref{fig_delta23}.  Because
of the two (three) fold symmetry for the $\epsilon_{2}$
($\epsilon_{3}$) moments, if the two angles are uncorrelated, the
distribution should be flat at one from 0.0 - $\pi/6$ radians. A clear
correlation is found between the two moments in peripheral Au+Au events. 
The correlation strength then decreases for more central events~\cite{alverendnote}. 


Previous studies have noted that correlations exist between the
positions of participating nucleons and are intrinsic to all Monte Carlo
Glauber calculations~\cite{Broniowski:2007}.  And at the same time, previous
publications assumed that this would not lead to correlations in the
vector direction of odd and event moments since the odd moments are
purely the result of fluctuations.

\begin{figure}[tbp]
  \centering
  \includegraphics[width=1.00\linewidth]{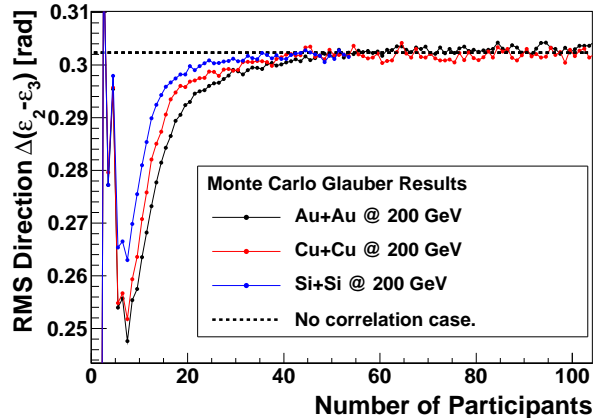}
  \caption{\label{fig_results23} 
(Color online) The root-mean-square (RMS) of the angular difference between $\epsilon_{2}$
and $\epsilon_{3}$ as a function of number of participant nucleons,
$N_{part}$, for Au+Au, Cu+Cu, and Si+Si collisions, as
well as the expectation of no correlation. We plot $N_{part} < 100$ to highlight
the significant correlations that are present at small $N_{part}$. 
}
\end{figure}

\begin{figure}[tbp]
  \centering
  \includegraphics[width=1.00\linewidth]{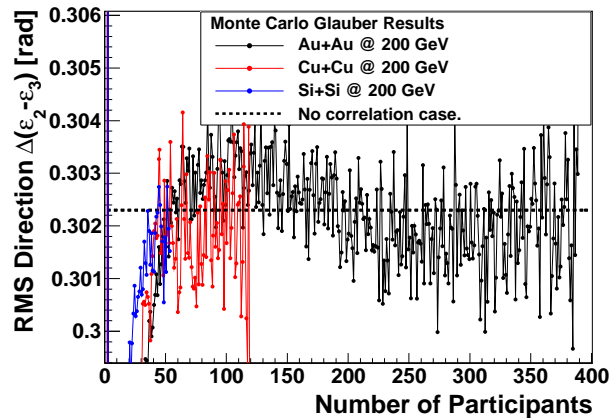}
  \caption{\label{fig_results23b} 
(Color online) The root-mean-square (RMS) of the angular difference between $\epsilon_{2}$
and $\epsilon_{3}$ as a function of number of participant nucleons,
$N_{part}$, for Au+Au, Cu+Cu, and Si+Si collisions, as
well as the expectation of no correlation. We zoom in on the vertical
axis to highlight the
small deviations from no correlation that remain at large $N_{part}$.
}
\end{figure}

\begin{figure}[tbp]
  \centering
  \includegraphics[width=1.00\linewidth]{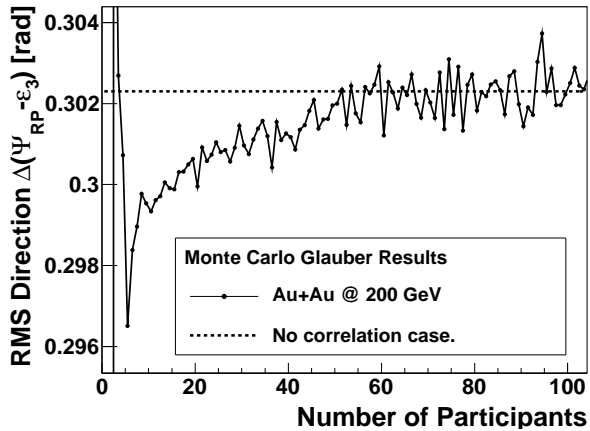}
  \caption{\label{fig_results3rp} 
(Color online) The RMS angular difference between the reaction plane, $\Psi_{RP}$, and $\epsilon_{3}$ as a
function of $N_{part}$ for Au+Au collisions. A weak
correlation is found between the 3rd-order moment and the reaction
plane direction for small $N_{part}$. Note the zoomed vertical scale.
}
\end{figure}

We quantify the degree of correlation by taking the root-mean-square 
(RMS) of the distribution. If there were no correlation between these
moments, the RMS of the angular separation would be 0.302 radians. The
results as a function of $N_{part}$ are shown in
Figure~\ref{fig_results23}. For our initial view, we have zoomed in
for $N_{part} < 100$. The results confirm the strong correlation
between the two angles (i.e. the large downward deviation in the RMS
from the flat case) for $N_{part} \approx 10$, the deviation then
weakens for larger $N_{part}$ values. In this figure, the correlation
appears to disappear for $N_{part} > 50$. However, in Figure~\ref{fig_results23b}, 
we show the full $N_{part}$ range
zooming the vertical axis around the default value. We observe a very
small remaining anti-correlation for $N_{part} \approx 80-150$, which
translates into an approximate 1\% lower probability of having the two
moments within 0.0-$\pi/12$ and a correspondingly higher probability
of having the two moments separated by $\pi/12$-$\pi/6$.   There is 
a similar magnitude positive correlation for $N_{part} > 200$.

\begin{figure}[t]
  \centering
  \includegraphics[width=1.00\linewidth]{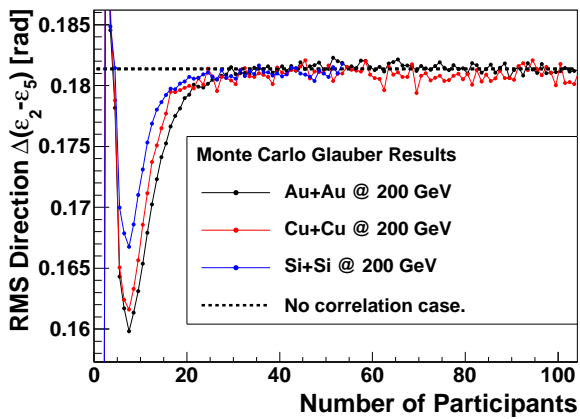}
  \caption{\label{fig_results25}
(Color online) The root-mean-square (RMS) of the angular difference between $\epsilon_{2}$
and $\epsilon_{5}$ as a function of number of participant nucleons,
$N_{part}$, for Au+Au, Cu+Cu, and Si+Si collisions, as
well as the expectation of no correlation. The result demonstrates
that a similar behavior holds for correlation between $\epsilon_2$ and
higher order odd moments. 
}
\end{figure}

\begin{figure}
  \centering
  \includegraphics[width=1.00\linewidth]{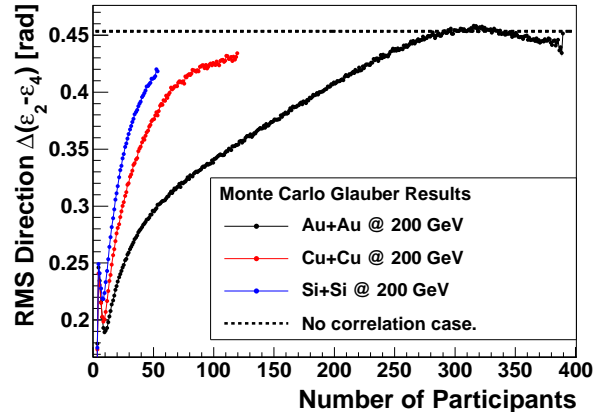}
  \caption{\label{fig_results24}
(Color online) The root-mean-square (RMS) of the angular difference between $\epsilon_{2}$ and
$\epsilon_{4}$ as a function of number of participant nucleons,
$N_{part}$, for Au+Au, Cu+Cu, and Si+Si collisions, as
well as the expectation of no correlation. The result shows the same procedure applied
to the difference between the lowest even-order moments where a
more significant correlation results from the average geometry.
}
\end{figure}

For the lower $N_{part}$ region where the alignment is strongest,
the correlation may be due to small number fluctuations of the
particular geometry in that impact parameter range for Au+Au. Thus,
also shown in Figure~\ref{fig_results23} are calculations from Si+Si
and Cu+Cu collisions. One can see that the angular
correlation largely tracks with $N_{part}$ and thus it is the number
fluctuations and not the average geometry that dominate the
correlation. However, the $Si, Cu, Au$ results do not scale
perfectly versus $N_{part}$, and thus the particular geometric
configuration plays some role, as one might expect since the geometry
correlates with the magnitude of these moments. 

Furthermore, the authors of~\cite{Petersen:2010cw} state that the fluctuations are
random with respect to the reaction plane (the plane defined the line
between the centers of the nuclei and the longitudinal axis). We have
tested this as well and show the resulting RMS angular separation between the
$\epsilon_{3}$ and the reaction plane in Figure~\ref{fig_results3rp}.
We find a similar trend as shown above. The correlation between
$\epsilon_3$ and the reaction plane is also restricted to small numbers of
participating nucleons.  The correlation here is much weaker
than what was found between $\epsilon_{2}$ and $\epsilon_{3}$. 

We also show the angular correlation between
$\epsilon_{2}$ and $\epsilon_{5}$ in Figure~\ref{fig_results25}.  One
again sees a strong correlation in peripheral collisions and then
smaller correlations for larger numbers of participants. Finally,
in Figure~\ref{fig_results24}, we show the angular correlation between
$\epsilon_{2}$ and $\epsilon_{4}$. These two even moments are expected
to be highly correlated because the initial overlap in mid-central
collisions has an approximately elliptical shape, which can be
well described by a combination of aligned $\epsilon_{2}$ and
$\epsilon_{4}$ moments if it has a large enough eccentricity. This 
correlation tracks with the geometry (i.e. impact parameter), and so
the resulting RMS values for Au+Au, Cu+Cu, and Si+Si do
not track each other when plotted as a function of $N_{part}$. 


In the case of nearly ideal hydrodynamics, the spatial eccentricities
may be translated into directly measurable momentum anisotropies ($v_{n}$).
The relative degree of correlation between $v_{2}$ and $v_{4}$ is of
great interest because it directly impacts a possible difference in
the observed $v_{4}$ determined via the event plane method (using the
second moment to determine the plane) and a two-particle Fourier
decomposition method~\cite{Luzum:2010sp}.  The $v_{3}$ moment is
identically zero by symmetry when measured with respect to the
second moment event plane.  However, the vector direction correlation
can be determined via event-by-event separate second and third-moment event
plane measurements.  Such measurements would provide potentially 
sensitive tests of the fluctuations in the Monte Carlo Glauber
geometry.  Despite the effect being largest in peripheral events where
nearly inviscid hydrodynamics may not apply, it should prove fruitful
to see if such angle correlations persist (which do not require a
linear translation of $\epsilon_{n}$ moments into $v_{n}$ momentum
anisotropies).

\section{Summary}

In summary, we find that contrary to previous publications, there is a
definite non-zero correlation within the Monte Carlo Glauber calculation between
the angular directions of $\epsilon_{2}$ and $\epsilon_{3}$ (and
more generally between even and odd moments). A much weaker
correlation between the reaction plane and $\epsilon_{3}$ is also
found. The effects are strongest for peripheral Au+Au events, and the
measurement of such correlations may provide an interesting test of
geometric fluctuations within the Monte Carlo Glauber calculation. 
It will be instructive to measure these possible correlations and care must be
taken to quote the exact sensitivity of such measures and extend them to the very most
peripheral events.
The multi-dimensional correlation between the magnitude
of the eccentricity orders and their angular orientations may also prove
important and the direct integration of Monte Carlo calculations of
initial state geometries may be warranted in many studies to
account for the full set of correlations.    

\begin{acknowledgments}
We thank B. Alver, P. Steinberg, and W. Zajc for valuable discussions. We
acknowledge funding from the Division of Nuclear Physics of the
U.S. Department of Energy under Grant No. DE-FG02-00ER41152. 
\end{acknowledgments}

\bibliographystyle{apsrev} \bibliography{geom_fluctuations}

\end{document}